\documentclass[prd,twocolumn,showpacs,nofootinbib,floatfix]{revtex4}

\ifx\pdfoutput\undefined
   \usepackage[dvips]{graphicx}
   \else
   \usepackage[pdftex]{graphicx}
   \fi
   \usepackage{epstopdf}

\usepackage{amsfonts}
\usepackage{amsmath}
\usepackage{amssymb}
\usepackage{bm}
\usepackage{dcolumn}
\usepackage{epsfig}
\usepackage{graphicx}
\usepackage{graphics}
\usepackage[latin1]{inputenc}
\usepackage{latexsym}
\usepackage{rotating}
\usepackage{hyperref}
\usepackage[usenames]{color}
\usepackage{float}
\newcommand\be{\begin{equation}}
\newcommand\ba{\begin{eqnarray}}
\newcommand\ee{\end{equation}}
\newcommand\ea{\end{eqnarray}}

\newcommand{\BD}{{\mbox{\tiny BD}}}

\newcommand{\GR}{{\mbox{\tiny GR}}}

\newcommand{\IM}{{\mbox{\tiny IM}}}

\newcommand{\LR}{{\mbox{\tiny LR}}}

\newcommand{\ISCO}{{\mbox{\tiny ISCO}}}

\newcommand{\SNR}{{\mbox{SNR}}}
\newcommand{\shift}{{\mbox{\tiny shift}}}
\newcommand{\fstretch}{{\mbox{\tiny stretch}}}

\begin{document}
\title{Mis-modelling in Gravitational Wave Astronomy: The Trouble with Templates}

\author{Laura Sampson}
\affiliation{Department of Physics, Montana State University, Bozeman, MT 59717, USA.}

\author{Neil Cornish}
\affiliation{Department of Physics, Montana State University, Bozeman, MT 59717, USA.}

\author{Nicol\'as Yunes}
\affiliation{Department of Physics, Montana State University, Bozeman, MT 59717, USA.}

\date{\today}

%%%%%%%%%%%%%%%%%%%%%%%%%%%%%%%%%%
%%%%%%%%%%%%%%%%%%%%%%%%%%%%%%%%%%
%%%%%%%%%%%%%%%%%%%%%%%%%%%%%%%%%%

\begin{abstract}

Waveform templates are a powerful tool for extracting and characterizing gravitational wave signals, acting as highly restrictive priors on the signal morphologies that allow us to extract weak events buried deep in the instrumental noise. 
The templates map the waveform shapes to physical parameters, thus allowing us to produce posterior probability distributions for these parameters.
However, there are attendant dangers in using highly restrictive signal priors. 
If strong field gravity is not accurately described by General Relativity (GR), then using GR templates may result in {\em fundamental} bias in the recovered parameters, or even worse, a complete failure to detect signals. 
Here we study such dangers, concentrating on three distinct possibilities. 
First, we show that there exist modified theories compatible with all existing observations that would fail to be detected by the LIGO/Virgo network using searches based on GR templates, but which would be detected using a one parameter post-Einsteinian extension. 
Second, we study modified theories that produce departures from GR that turn on suddenly at a critical frequency, producing waveforms that do not directly fit into the simplest parameterized post-Einsteinian (ppE) scheme. 
We show that even the simplest ppE templates are still capable of picking up these strange signals and diagnosing a departure from GR.
Third, we study whether using inspiral-only ppE waveforms for signals that include merger and ringdown can lead to problems in misidentifying a GR departure. 
We present a simple technique that allows us to self-consistently identify the inspiral portion of the signal, and thus remove these potential biases, allowing GR tests to be performed on higher mass signals that merge within the detector band. 
We close by studying a parameterized waveform model that may allow us to test GR using the full inspiral-merger-ringdown signal.
\end{abstract}

%%%%%%%%%%%%%%%%%%%%%%%%%%%%%%%%%%
%%%%%%%%%%%%%%%%%%%%%%%%%%%%%%%%%%
%%%%%%%%%%%%%%%%%%%%%%%%%%%%%%%%%%

\maketitle

%%%%%%%%%%%%%%%%%%%%%%%%%%%%%%%%%%
%%%%%%%%%%%%%%%%%%%%%%%%%%%%%%%%%%
%%%%%%%%%%%%%%%%%%%%%%%%%%%%%%%%%%
\section{Introduction}

Gravitational waves (GWs) carry vast amounts of energy, but are notoriously difficult to detect because of their weak coupling to matter. The first direct detections will likely come from
signals buried in the instrument noise that can only be extracted using sophisticated data analysis techniques. If the form of the signal, $h(t)$, can be predicted in advance, we
can use this information as a prior. This can greatly aid in separating signals from noise in the data, $s(t)$, by demanding that the residuals, $r(t) = s(t)-h(t)$, are consistent with
our instrument noise model. The different classes of analyses can be classified by the strength of the priors~\cite{Cornish:2013nma}, ranging from the weak signal priors used in
burst searches~\cite{Sutton:2013ooa, East:2012xq,Abadie:2012rq,Was:2012zq,Mohapatra:2011mx, Abadie:2010wx,Abadie:2010mt} and
stochastic searches~\cite{Adams:2013qma,Nishizawa:2013eqa,Mingarelli:2013dsa,Tinto:2012ba,Demorest:2012bv,Thrane:2013kb,Mandic:2012pj}, to the highly restrictive priors used in searches for binary mergers~\cite{Colaboration:2011np,Brown:2012nn,Farr:2009pg,VanDenBroeck:2009gd,Abadie:2010uf,Abadie:2011kd,Ajith:2012az,Aasi:2012rja}.  

When considering signals from binary systems, the signal prior is strong because we believe that we can model the waveforms very accurately. Waveform models for binary systems map the signals to
system parameters, resulting in a template $h(t) \rightarrow h(\vec{\lambda})$, which allows the detection of signals that are orders of magnitude weaker than the instrument noise level.
These templates allow us to measure certain physical parameters of the astrophysical systems that generate GWs, like the masses and spins of the bodies in a binary~\cite{Rodriguez:2013oaa,O'Shaughnessy:2013vma,Littenberg:2013gja,Aasi:2013jjl,Lackey:2013axa}. Template-based approaches can also be used to test the accuracy of GR~\cite{Cornish:2011ys,Sampson:2013lpa,Chatziioannou:2012rf,Sampson:2013wia,Yunes:2009ke,Li:2011cg,Agathos:2013oma,Mishra:2010tp,Huwyler:2011iq,PhysRevD.86.044010} (see~\cite{Yunes:2013dva} for a recent review of GW tests of GR with ground-based instruments). 

However, the power of a template based search has attendant dangers: when the prior is far more informative than the likelihood, it dominates the posterior probability distribution. This can lead to large systematic biases if the model is an imperfect description of Nature~\cite{Yunes:2009ke,Yagi:2013baa,Favata:2013rwa}. When the disagreement between the model and reality is too large, a template
based search will fail to detect any signals in the data. 

Analyses of the data collected by the first generation of ground based interferometers have failed to detect any signals~\cite{Colaboration:2011np,Abadie:2010uf,Aasi:2013jya,TheLIGOScientific:2013cya,Aasi:2013sia,Aasi:2012fw,Abadie:2012rq}. Could this be because we have been looking in the wrong
place? While we do not consider this possibility to be very likely, we show that existing observational bounds on alternative theories of gravity do not preclude the possibility.
There are a wide class of theories that agree with GR in the slow motion, weak field regime, and pass all existing tests, yet differ significantly in the strong field, dynamical
regime probed by binary inspiral and merger, e.g.~dynamical Chern-Simons gravity~\cite{Alexander:2007vt,Smith:2007jm,Alexander:2009tp,Yunes:2009hc,Sopuerta:2009iy,AliHaimoud:2011fw,Yagi:2011xp,Yagi:2012ya,Yagi:2012vf,Yagi:2013mbt}. We show that GWs from these types of theories could be missed entirely by searches using GR templates. Higher mass
mergers and other bright signals that stand up above the instrument noise would be detected by the less restrictive burst search techniques, so it is unlikely that the lack
of detections points to a significant departure from GR, but the possibility remains. Here we show that a simple one-parameter extension to the usual GR templates restores
sufficient flexibility to detect most departures from GR in inspiral signals.

The second issue we explore is whether we can use non-GR templates designed with a certain class of theory in mind to detect deviations from GR that arise from an entirely different class of modification. In particular, we determine whether a family of non-GR templates, called the parameterized post-Einsteinian (ppE) family, can be used to detect non-GR signals that they were not designed to capture. The ppE scheme has previously been shown to map to the predicted GW signals from binary inspirals in all known, analytic alternative theories of gravity. In this section, we show that the ppE templates can also encapsulate the changes to GW signals that would arise from violations of the no-hair theorem for black holes. 

Later in the paper, we examine non-GR, non ppE-like, signals that are predicted from certain massive theories~\cite{Alsing:2011er,Cardoso:2011xi,Berti:2012bp} or certain scalar-tensor (ST) theories~\cite{Barausse:2012da,Shibata:2013pra,Palenzuela:2013hsa}, in which the non-GR part of the GW signal ``turns-on'' at a critical frequency, $f^*$, and is not present at frequencies lower than this. We find that the standard ppE templates work well for detecting this type of deviation in some cases, and that a simple modification to the ppE templates works well in all cases.

Next, we examine the issue of using inspiral-only templates to characterize signals that consist of inspiral, merger, and ringdown. Because we lack knowledge of the form of merger and ringdown in modified theories, we commonly choose to use only the inspiral portion of the waveform in analyses that seek to test GR. Deciding which portion of a GW signal is the inspiral, however, is not a trivial problem, and different choices of how to make this distinction lead to different results in our analysis. We find that, for the low-mass systems, inspiral-only searches do not lead to biased parameter estimations. But for higher-mass systems it can lead to large biases in recovered parameters, and even the false claim of a detection of a deviation from GR. We show that there is a simple technique for avoiding these biased results, which allows us to use some high-mass systems in inspiral-only studies to test GR. 

Finally, we refine the ppE parameterization for the merger and ringdown of binary systems~\cite{Yunes:2009ke}, and examine how it could be used to test GR with full signals. Lacking concrete examples of merger/ringdown in alternative theories, our parameterization is fairly arbitrary. It does, however, capture a wide range of deviations from GR that may arise in this sector. We find that for systems with total mass $M\gtrsim 50 M_{\odot}$, this parameterization could be used to learn about deviations from GR.

The remainder of this paper presents the details of the results discussed above. 
Section~\ref{sec:GW-in-alt-theories} introduces the ppE parameterization in more detail.
Section~\ref{sec:detecting-non-GR} shows that certain modified gravity signals that are not ruled out by present observations could be missed by matched filtering searches with GR templates.  
Section~\ref{sec:detect-ST} studies the effectiveness of ppE templates to detect certain modified gravity signals that turn on at a critical frequency. 
Section~\ref{sec:ins-analysis} studies whether inspiral ppE templates can be used to analyze high-mass signals that include merger and ringdown.
Section~\ref{sec:Conc} concludes and points to future research. 

Throughout this paper we use geometric units, in which $G=c=1$. We perform all analyses using an Adv.~LIGO (aLIGO) noise curve that is optimized for detection of inspiraling neutron stars (see e.g.~\cite{website:aLIGOcurves}, the curve titled NSNS\_Opt).

\section{The Parameterized post-Einsteinian Framework}
\label{sec:GW-in-alt-theories}

In order to study the pitfalls of template-based analyses in detecting signals and testing GR, we need a model for what gravitational wave signals might look like in alternative theories of gravity.
The ppE template family can represent the signals produced by binary systems in a large class of alternative theories. Developed by Yunes and Pretorious~\cite{Yunes:2009ke}, and explored in~\cite{Sampson:2013lpa,Cornish:2011ys}, these templates for the Fourier transform of the quadrupole GW strain signal from a system of two inspiraling, non-spinning, compact objects in quasi-circular orbits take the form
\begin{align}
\tilde{h} (f) = \tilde{h}^{\GR}\cdot (1&+\sum_a\alpha_{a}  u^{a}) \exp{i \sum_b \beta_{b} u^{b}},\\ &u = (\pi \mathcal{M} f)^{1/3},
\label{eq:ppEtemp}
\end{align}
where $\alpha_a$ and $\beta_b$ are the ppE strength parameters associated with the exponents $a$ and $b$, $\mathcal{M} = (m_1 m_2)^{3/5}/(m_1+m_2)^{1/5}$ is the chirp mass of the system, and $\tilde{h}^\GR$ is the usual GR expression for a non-spinning, inspiraling binary~\cite{Blanchet:2013haa}. The extension of this template family to include other polarizations has been considered in Ref.~\cite{Chatziioannou:2012rf}.

As mentioned, these templates can match the gravitational waves generated in a wide class of alternative theories of gravity. The exact correspondence between the ppE parameters and a number of alternative theories has been discussed in~\cite{Cornish:2011ys,Sampson:2013wia,Yunes:2013dva}. What has not been discussed before, though, is the fact that the ppE templates can also capture deviations from GR that would arise from a violation of the no-hair theorem for black holes. The derivation of this correspondence can be seen in~\cite{PhysRevD.56.1845}, in which Ryan calculates the GW signal that would result from such a violation. 

Assuming a GW signal is generated by two slowly inspiraling, compact objects whose exterior gravitational field can be parametrized by a set of arbitrary multipole moments, Ryan shows that the resulting GW phase can be written
\begin{multline}
\psi(f) =  \psi_\GR + \frac{3}{128 \eta} \sum_{\mbox{odd} \,\,\ell \ge 3}\frac{(-1)^{\ell-1}80(2\ell+5)!! s_\ell u^{2\ell-4}}{3(\ell-2)(2\ell-7)(\ell-1)!!} \\
                -\frac{3}{128\eta}\sum_{\mbox{even}\,\, \ell\ne 4}\frac{(-1)^{\ell/2}40(2\ell+1)(\ell+1)!!m_\ell u^{2\ell-5}}{3(2\ell-5)(\ell-4)!!},
\end{multline}
where $m_\ell$ and $s_\ell$ are the (dimensionless) $\ell^{th}$ mass and current multipoles, respectively, and high-order terms have been dropped. This result can clearly be mapped to the ppE templates with the substitution
\be
b_\ell \rightarrow \begin{cases} 2\ell -4, \,\,\, \mbox{current multipole}, \\ 2\ell -5, \,\,\, \mbox{mass multipole},
\end{cases}
\ee

\be
\beta_\ell \rightarrow \begin{cases}\frac{3}{128 \eta} \frac{(-1)^{\ell-1}80(2\ell+5)!! s_\ell}{3(\ell-2)(2\ell-7)(\ell-1)!!} , \,\,\, \mbox{current multipole}, \\ -\frac{3}{128\eta}\frac{(-1)^{\ell/2}40(2\ell+1)(\ell+1)!!m_\ell }{3(2\ell-5)(\ell-4)!!}, \,\,\, \mbox{mass multipole}.
\end{cases}
\ee

The largest deviation in the phase corresponds to the lowest-$\ell$ multipoles. For example, a modification to the mass quadrupole ($\ell=2$) induces a 2PN correction to the waveform that can be captured by a $b=-1$ ppE modification.  The next largest modification enters through the current octopole ($\ell=3$), which induces a 3.5PN correction that can be captured by a $b=2$ ppE modification. This octopole term is already at high PN order, and so higher order terms than this would be difficult to detect given a signal with reasonable SNR.

The ppE  templates are derived by introducing parameterized deviations to the orbital binding energy and GW luminosity of GR, and then calculating the resultant GW signal via a standard procedure~\cite{Blanchet:2013haa}. Although in theory this type of modification introduces an infinite number of non-GR terms in both the phase and the amplitude of GWs, it was shown in~\cite{Cornish:2011ys} that modifications to the GR amplitude are poorly constrained by GW measurements. Additionally, in~\cite{Sampson:2013lpa} it was shown that simple ppE templates of the form in Eq.~(\ref{eq:ppEtemp}), but in which the sum over phase modifications is replaced by a single term, perform well in detecting full ppE signals. That is, even when the injected signals contain many modifications to the GR expressions, a ppE template containing only one modification is sufficient for extracting the signal.

For these reasons, in this paper, \emph{when recovering signals}, we will focus on using ppE inspiral templates of the form
\be
\tilde{h} (f) = \tilde{h}^{\GR}\cdot  e^{i \beta_{b} u^{b}},
\label{eq:ppEtemp_simple}
\ee
which contain only a single modification to the GR phase, and no modifications to the amplitude. Note, though, that we will include more complicated signals, as explored in~\cite{Sampson:2013lpa}, for signal \emph{injections}. We stress that this is the \emph{simplest} ppE model conceivable; more complicated generalizations that include the merger and ringdown~\cite{Yunes:2009ke} and additional polarizations~\cite{Chatziioannou:2012rf} exist, but are not needed for the present analysis. 
 
\section{Missing non-GR signals with GR templates}
\label{sec:detecting-non-GR}
Different non-GR theories of gravity lead to modifications to the phase and amplitude of GWs at different post-Newtonian (PN) orders. The ppE parameters can, of course, be mapped to these PN orders. We show this mapping in Table~\ref{PNorders}.
\begin{table}[ht]
\begin{center}
    \begin{tabular}{c|c|c }
    \hline\hline
    b & PN order& Physical effects \\ 
    \hline
    $-7$&$-1$ PN&Dipole radiation\\
    $-5$&$0$ PN&Quadrupole radiation\\
    $-3$&$1$ PN&Mass ratio measured\\
    $-2$&$1.5$ PN&Spin effects enter\\
    $-1$&$2$ PN&--\\
    $0$&$2.5$ PN&--\\
    $1$&$3$ PN&--\\
  \hline\hline
           \end{tabular}
   
    \caption{ \label{PNorders}Correspondence between ppE $b$ parameters and PN order, as well as selected physical effects related to each order.}
\end{center}
\end{table}

\begin{figure}
\begin{center}
\begin{tabular}{cc}
\hspace*{-8mm} \includegraphics[width =8.5 cm]{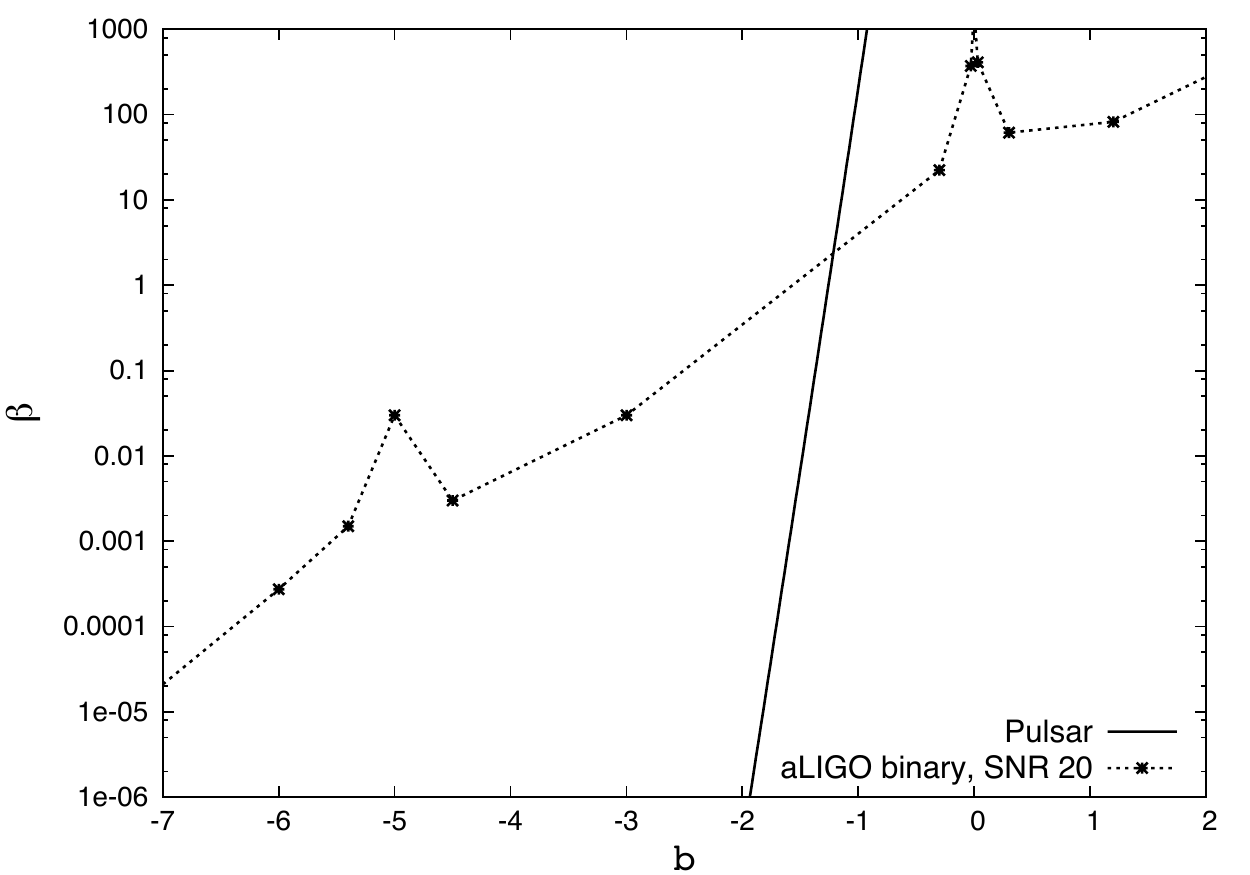} 
\end{tabular}
\end{center}
\vspace*{-0.2in}
\caption{\label{bbetabounds} Bounds that can be placed on the ppE strength parameter, $\beta$, for various values of the ppE exponent, $b$, using GW measurements~\cite{Cornish:2011ys} from a binary with component masses of $6 M_\odot$ and $12 M_\odot$, and measurements from binary pulsars~\cite{Yunes:2010qb}. The regions above the pulsar line are already ruled out by experiment. Ref.~\cite{Cornish:2011ys} shows that this bound is robust over a range of mass ratios.}
\end{figure}

 For theories that lead to low PN-order\footnote{A modification at NPN order is one which is proportional to $(v/c)^{2N}$ relative to the leading-order term, where $v$ is the characteristic orbital velocity of the binary and $c$ is the speed of light.} deviations from GR, e.g.~Brans-Dicke gravity leads to changes in the phase at $-1$PN order relative to the leading-order GR prediction~\cite{Yunes:2009ke}, there are already strong constraints on the possible size of the deviation from current Solar System and pulsar timing experiments. The current bounds placed on ppE parameters by pulsar timing experiments are shown in Fig.~\ref{bbetabounds}~\cite{Yunes:2010qb}. As stated, for large, negative values of $b$, i.e.~low-PN order terms, current data from pulsar binary systems places very tight bounds on possible deviations in the phase of GW signals from the GR expectation. 

On the other hand, for theories that lead to high PN-order deviations, the existing constraints are very weak. This is also shown in Fig.~\ref{bbetabounds}, which shows that at high-PN order, i.e. for less negative values of $b$, the bounds become very weak. Therefore, it is possible that GW signals will differ greatly from the GR predictions, but only once the characteristic velocity of the system becomes quite high, thus avoiding current experimental bounds. In this section, we examine the possible effects of using GR templates to search for non-GR signals that contain large deviations from GR at high PN order. Although `large' in the sense that they would be easily differentiable from GR, these deviations are not currently ruled out by any experimental evidence.

To test what effect a signal containing these high-order PN deviations could have on our ability to detect and characterize GWs using GR templates, we inject non-spinning ppE inspiral signals of the form of Eq.~(\ref{eq:ppEtemp}), containing two phase corrections, with $b=-1$ and $b=1$. 
 This corresponds to adding both a 2PN  and a 3PN order correction to the GW phase. The $\beta_{-1}$ parameter is chosen to be a deviation from GR that is not ruled out by current experimental bounds, and $\beta_{1}$ is chosen to be larger by a factor of 1000. 
 That is,
\be
\tilde{h}_{\rm inj} = \tilde{h}_{\GR}\left(\exp{i \left[\beta_{-1} u^{-1}+ 1000 \times \beta_{-1} u^{1}\right]} \right).
\label{nonGRtwoterms}
\ee
 This factor is a conservative one - based on the bounds on $\beta_b$ shown in Fig. \ref{bbetabounds}, it is clear that an even larger ratio would be well within experimental limits. The system studied is a $1.4 M_\cdot$ neutron star and a $3.5 M_\odot$ black hole in a quasi-circular orbit. The effect of using non-spinning templates for black holes in this type of study was examined in \cite{Sampson:2013lpa}, and was found to be minimal.

Having injected this non-GR signal into our data, we then calculate the fitting factor (FF),
\be
{\rm FF} = \max\Bigg[\frac{(h|s)}{\sqrt{(h|h)}\sqrt{(s|s)}} \Bigg].
\ee
In this expression, $(a|b)$ is the noise-weighted inner product between $a(f)$ and $b(f)$, integrated starting at 2 Hz, $s$ is the actual signal in our detector, and $h$ is the template being used to recover this signal. (All calculations throughout this paper used a minimum frequency threshold of 2Hz.) The FF is maximized over system parameters, which, for a circular binary, are total mass $M$, chirp mass $\mathcal{M}$, luminosity distance $D_L$, phase and time of coalescence, $\phi_c$ and $t_c$, two sky location angles, $\alpha$ and $\delta$, and the inclination angle, $\iota$. Clearly, if the template is precisely the same as the signal, ${\rm FF}=1$. The FF is equal to the fraction of the signal-to-noise ratio (SNR) that we can recover with our templates:
\be
\SNR_{\rm rec} = {\rm FF} \times \SNR,
\ee
where $\SNR_{\rm rec}$ is the recovered SNR, while $\SNR$ is the true SNR if we had exactly the right template. The fractional loss of events due to mis-modeling errors in the templates scales as $1-{\rm FF}^3$, so to achieve $90\%$ efficiency
we must have ${\rm FF} > 0.97$. FFs below $0.5$ imply a 90\% loss of signals, while FFs below 0.2 imply a 99\% loss of signals.
Thus,  if the  FF of GR templates with plausible non-GR signals is significantly lower than $0.97$, we could miss GW signals that are present in our data streams. 

\begin{figure}
\begin{center}
\vspace*{-0.22in}
\begin{tabular}{cc}
\hspace*{-8mm} \includegraphics[width =9.5 cm]{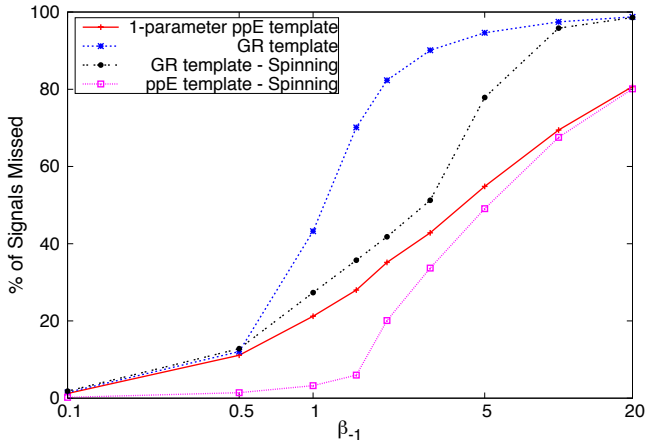} 
\end{tabular}
\end{center}
\vspace*{-0.42in}
\caption{\label{FF2loss}  Percentage of signals missed ($100\times(1-FF^3)$) when using  a GR (dashed/blue line) and a one-parameter ppE template with $b=-1$ (solid/red line) and the injected, non-GR signal (of the form in Eq.~\ref{nonGRtwoterms}, with constituent masses $1.4 M_\odot$ and $3.73 M_\odot$) as a function of $\beta_{-1}$. ($b=-1$ corresponds to a 2PN correction to the GR waveform.) The same quantity is plotted for spinning GR and ppE templates. The percentage
of signals missed approaches $100\%$ for values of $\beta_{-1}$ that
are fully consistent with existing experimental bounds.}
\end{figure}

\begin{figure}
\begin{center}
\begin{tabular}{cc}
\hspace*{-8mm} \includegraphics[width =9.5 cm]{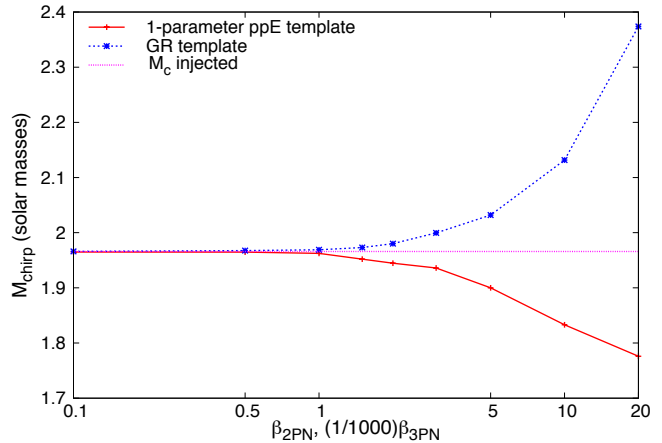} 
\end{tabular}
\end{center}
\vspace*{-0.4in}
\caption{\label{deltaMchirpFF} Recovered value of $\mathcal{M}$ from both a GR (dashed/blue line) and a one-parameter ppE template with $b=-1$ (solid/red line) used to fit a non-GR signal of the form in Eq.~\ref{nonGRtwoterms}, plotted as a function of $\beta_{-1}$. The true value of $\mathcal{M}$ is $\mathcal{M} = 1.94 M_\odot$. The constituent masses were $1.4 M_\odot$ and $3.73 M_\odot$. Using GR templates to fit non-GR signals leads to large biases in the recovered parameters.}
\end{figure}

Figure~\ref{FF2loss} shows the percentage of signals that would be missed by fitting our injected, non-GR signals with non-spinning, circular GR templates. For small values of $\beta_{-1}$, as expected, the GR templates can achieve a near-perfect FF. Quite quickly, however, as $\beta_{-1}$ increases, the FF drops below the desired level of ${\rm FF}=0.97$. Well before the injected signals are ruled out by current pulsar constraints, the FF drops to $0.2$,
and it is likely that all such signals would be missed. The decrease in FF is accompanied by an increasing bias in the recovered chirp mass, shown in Fig.~\ref{deltaMchirpFF}. The prior range in chirp mass was from 
$0.87$ to $8.71$ $M_\odot$, which corresponds to a range in the individual masses from $1$ to $10$ $M_\odot$. It is clear that the recovered values for $\mathcal{M}$ are still well within the prior range, and so increasing this will not increase the FFs.

\begin{figure}
\begin{center}
\begin{tabular}{cc}
\hspace*{-8mm} \includegraphics[width =9.5 cm]{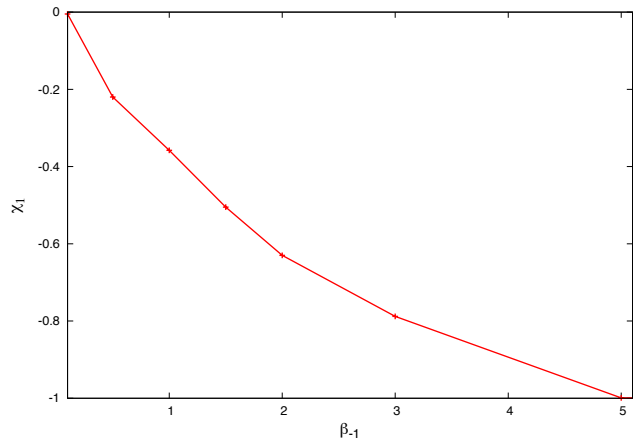} 
\end{tabular}
\end{center}
\vspace*{-0.4in}
\caption{\label{Spin} Recovered value of $\chi_1$ from a spinning, GR template used to match a non-spinning, non-GR signal of the form in Eq.~\ref{nonGRtwoterms}. Once $\beta_{-1} \gtrsim 5$, $\chi_1$ is limited by the prior range.}
\end{figure}

Also shown in Figure~\ref{FF2loss} is the percentage of signals missed when fitting the injections using circular GR templates that include aligned spin. This introduces two new parameters, the dimensionless spin parameters. $\chi_1$ and $\chi_2$, which are by definition in the prior range $-1\le\chi_A\le1$. The extra freedom from including these parameters allows the spinning GR templates to achieve a better fit than the non-spinning templates for some injected values of $\beta_{-1}$. However, once $\beta_{-1}$ is large enough, the spinning template performs no better than the non-spinning. This can be understood by examine Figure~\ref{Spin}. This figure shows the recovered value for $\chi_1$ from the spinning GR templates (recall that the injected signals were \emph{non-spinning}). It is clear that after $\beta_{-1} \sim 5$, the spin parameter has reached the limit of the prior range, and thus has reached the limit of its ability to help with the fit. After this point, the FFs for spinning and non-spinning GR templates converge.

Figures~\ref{FF2loss} and~\ref{deltaMchirpFF} show the percentage of signals missed and recovered chirp mass for an analysis using a simple ppE template that contains only one strength parameter, $\beta_{-1}$, with exponent $b=-1$ (recall that the injected signals have two ppE strength parameters -$\beta_{-1}$ and $\beta_{1}$). These templates perform much better than the GR ones at detecting the signal, but suffer from similar issues in biased recovery of the chirp mass. Finally, Figure~\ref{FF2loss} shows the increase in detection efficiency that can be achieved by using a one-parameter ppE template that also has the two aligned spin parameters, $\chi_1$ and $\chi_2$. As expected, this template family, which contains the most free parameters, performs the best at detecting the injected signal.

This outcome is not surprising -- we know that templates are only effective in detecting signals that are at least somewhat similar to them. It is important, though, to be aware that there are non-GR signals, completely consistent with current experiment, that would be entirely missed using GR templates.

\begin{figure}
\begin{center}
\begin{tabular}{cc}
\hspace*{-8mm} \includegraphics[width =9.5 cm]{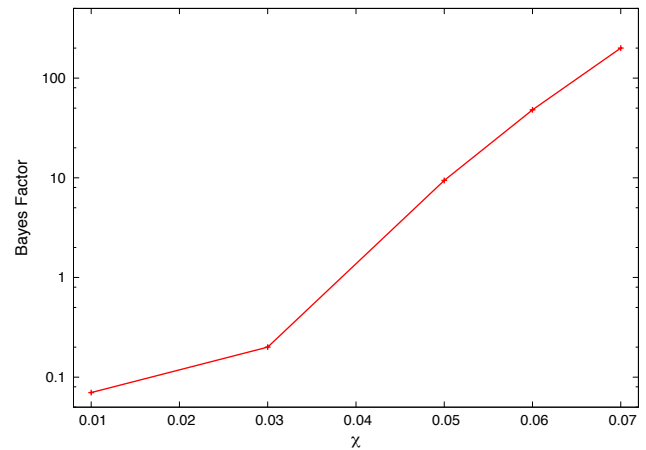} 
\end{tabular}
\end{center}
\vspace*{-0.4in}
\caption{\label{BFspins} The BFs between a non-spinning, GR template and a non-spinning, ppE template with $b=-2$.The injection was a GR system of two neutron stars with equal and aligned spin. The BF favors the non-GR model for realistic values of dimensionless spin parameter $\chi$, indicating the need to use spinning waveforms to recover these types of signals. }
\end{figure}

The result shown in Figure~\ref{FF2loss} indicates that spin can play a major role
in tests of GR, and raises the question of when spin effects may be
masquerade as a departure from GR when not properly accounted for
in the analysis. Fugure~\ref{BFspins} shows the Bayes Factors between a GR model and
a one parameter $b=-2$ ppE model that both neglect spin effects. The Bayes Factors
are shown as a function of the total spin parameter of the injected signals for
a binary NS system with $m1=m2 = 1.4M_\cdot$ and $\mbox{SNR} = 12$. The additional ppE
phase term is able to mimic the leading order spin effect, leading
to a clear preference for the alternative gravity model for spins above 0.05.
Clearly, spin effects need to be taken into account even at the small spin
values expected for NS binaries.

\section{Detecting GWs from Scalar-Tensor Gravity}
\label{sec:detect-ST}
The ppE waveforms cover all known inspiral waveforms from specific alternative theories of gravity that are analytic in the frequency evolution of the GWs~\cite{Cornish:2011ys}. There are, however, known theories for which the frequency evolution of GWs is \emph{not} analytic -- in particular, ST theories of gravity in which spontaneous scalarization can occur, and theories of gravity that contain a massive scalar field~\cite{Alsing:2011er,Cardoso:2011xi,Barausse:2012da, Shibata:2013pra}.

The theories that include spontaneous scalarization can be defined with a generic ST action of the form~\cite{Barausse:2012da, Shibata:2013pra}
\be
S = \int d^4 x \frac{\sqrt{-g}}{2\kappa} \Bigg[\phi R - \frac{\omega(\phi)}{\phi} \partial_\mu \phi \partial^\mu \phi \Bigg] + S_M .
\label{eq:staction}
\ee
where $\kappa = 8\pi G$, $R$ is the Ricci scalar, $g$ is the determinant of the metric, $\phi$ is the gravitational scalar, and $S_M$ is the matter action.

In this type of theory, neutron stars that are not initially scalarized can acquire a scalar charge when the system reaches a high enough binding energy -- i.e.~once the orbital frequency is high enough. This spontaneous change leads to the ``turning on'' of dipole GW radiation once the merging stars get close enough together. This radiation, in the ppE scheme, goes as $b=-7$, a low PN order effect -- lower order, in fact, than even the Newtonian term. This means that binary pulsar measurements have placed very tight restrictions on the possible strength of this dipole radiation at low frequencies~\cite{Antoniadis:2013pzd,Freire:2012mg}. However, we do not yet have any measurements of binary systems at high frequencies. It is therefore not impossible that signals of this type, that have no dipole radiation at low frequencies, but significant dipole radiation at high frequencies, could be detected by GW experiments.

Another type of ST theory that produces this type of radiation is that in which the scalar field, $\phi$, has a mass~\cite{Alsing:2011er,Cardoso:2011xi}. The gravitational radiation due to this type of theory has been calculated in~\cite{Berti:2012bp}. The phase is altered from the GR expression, and is equal to
\begin{align}
\psi(f) = \psi_\GR(f) & + \xi \Gamma^2 \nu \Bigg[\frac{5}{462} u^{-11} -\frac{\nu}{1632} \eta^{12/5}u^{-17} \Bigg] \nonumber\\ &\times\Theta(2\pi f - m_s)
\nonumber\\ & + \xi \mathcal{S}^2 \Bigg[\frac{25 \nu}{1248} \eta^{8/5} u^{-13} - \frac{5}{84} \eta^{2/5} u^{-7} \Bigg] \nonumber\\ & \times \Theta(\pi f - m_s).
\label{eq:theta phase}
\end{align}
In this expression, $m_s$ is the mass of the scalar field, $\eta = (m_1 m_2)^{3/5}/(m_1+m_2)^{1/5}$ is the symmetric mass ratio and the other quantities are given by
\begin{align}
\xi  &= \frac{1}{2+\omega_\BD}, \nonumber \\
\Gamma &= 1 - 2\frac{s_1m_2+s_2m_1}{M}, \nonumber \\
\mathcal{S}& = s_2-s_1, \nonumber \\
\nu &=  5.60 \times 10^{-21}\Big(\frac{m_s}{10^{-20} eV} \frac{M}{M_{\odot}} \Big),
\end{align}
where $s_1$ and $m_1$ are the sensitivity and mass of body 1, $M$ is the total mass of the system , and $\omega_\BD$ has been constrained by the Cassini spacecraft such that $\omega_{\BD} \ge 40000$. The sensitivity is defined as
\be
s_A = \frac{d \ln M_A(\phi)}{d \ln \phi}.
\ee
The non-GR portions of the signal are not present until a frequency $f = m_s/2\pi$. This means that, in order for the non-GR signal to be detectable in the aLIGO band, i.e.~$f\sim100$ Hz, the mass of the scalar field must be approximately $m_s \sim 10^{-13 } \; {\rm{eV}}$. This leads to a value of $\nu \sim 10^{-14}$, which implies that the phase terms that are multiplied by $\nu$ are highly suppressed in this frequency range, leaving only dipole deviations from GR.

Thus both of these types of ST GW signals can be approximated, in ppE notation, as
\be
\tilde{h}(f) = \tilde{h}_\GR e^{i \Theta(f - f^*) \beta u^{-7}}.
\label{eq:phasetheta}
\ee
Here, $\Theta(f-f^*)$ is the Heaviside function, and $f^*$ is the frequency at which the dipole radiation `turns on.' There is an obvious discontinuity at $f=f^*$, which captures the discontinuity inherent in signals from these types of theories. Clearly, as $f^*$ goes to infinity, the signal in Eq.~(\ref{eq:phasetheta}) becomes a GR signal, and when $f^*$ is only a few Hz, the signal is vastly different from GR. 

\begin{figure}[t]
\begin{center}
\begin{tabular}{cc}
\hspace*{-8mm} \includegraphics[width =9.5 cm]{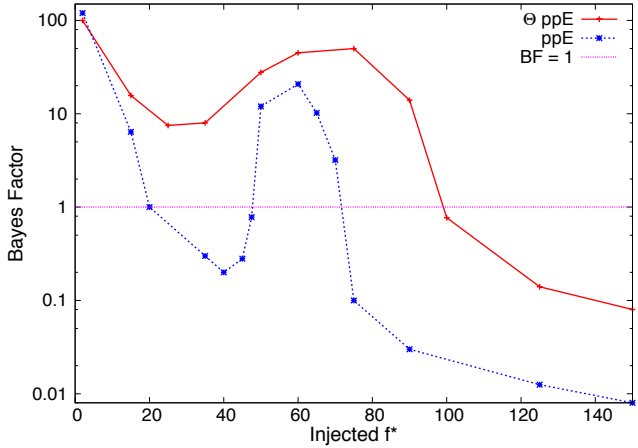} 
\end{tabular}
\end{center}
\vspace*{-0.4in}
\caption{\label{BFtheta} BF between GR and ppE models for injections of the form in Eq.~(\ref{eq:phasetheta}), for varying values of $f^*$. The dashed (blue) line is for a standard and simplest, inspiral ppE template, and the solid (red) line is one that has been modified with a Heaviside function. Both types of templates show the same general behavior, and both are successful at detecting deviations from GR for certain ranges of $f^*$. The injected signal had constituent masses  $m_1/m_2 = 1.42/2.0 M_\odot$.}
\end{figure}

Our first task, then, is to determine at what value of $f^*$ the standard ppE templates, with $b=-7$, will detect the non-GR character of the signal. To find this value, we inject signals of the form in Eq.~(\ref{eq:phasetheta}) with $\beta= 10^{-6}$, and varying values of $f^*$, ranging from 5 Hz to 150 Hz. These signals are SNR~12 inspirals from neutron star binary systems with constiuent masses of $1.4$ and $2.0$ $M_\odot$, using the neutron-star-binary optimized aLIGO noise curve. We then use Markov Chain-Monte Carlo (MCMC) techniques to recover the signals and calculate the Bayes factor (BF) between GR and non-GR models. The BF between two models, A and B, is the `betting odds' that model A provides a better description of the data than model B. If the BF of A vs.~B is greater than one, the data shows a preference for model A. In this paper, we compute BFs using the Savage-Dicke density ratio~\cite{1995:Savage}, in which the prior weight at $\beta=0$ is compared to the posterior weight. A decrease in probability density at this point corresponds to a preference for a non-GR model. The BFs between GR and non-GR are plotted in Fig.~\ref{BFtheta}. When the BF is above $1$, in an ideal study, the model selection process favors a non-GR signal. In reality, in order to claim a detection of deviation from GR, a BF much greater than $1$ would be required. 

\begin{figure}
\begin{center}
\begin{tabular}{cc}
\hspace*{-8mm} \includegraphics[width =9.5 cm]{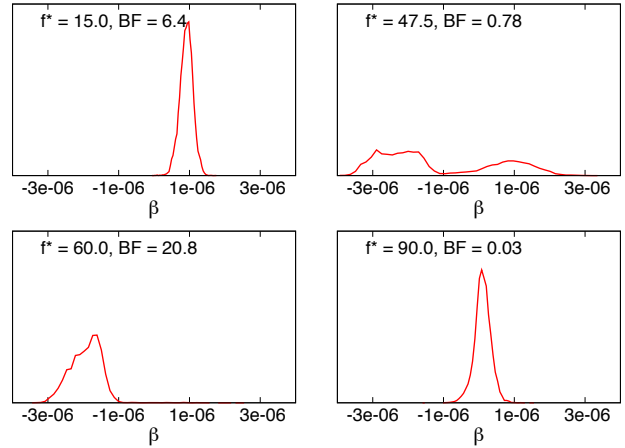} 
\end{tabular}
\end{center}
\vspace*{-0.4in}
\caption{\label{fourPDF} Posterior distributions for $\beta$, recovered using standard ppE templates. The injected signal was of the form in Eq.~(\ref{eq:phasetheta}), with $\beta =$ 1e-06, and $m_1/m_2 = 1.42/2.0 M_\odot$. If there is little weight in the posterior at $\beta=0$, the signal is detectable as non-GR. In the top left panel, $f_{min}$ is low, and $\beta$ is recovered at the correct value. In the bottom right panel, $f_{min}$ is very high, and the GR model is clearly favored. In the bottom right panel, the signal is clearly non-GR, but the recovered value for $\beta$ is incorrect. Finally, in the top right panel, two peaks in the posterior are clearly visible -- one mode near the correct value of $\beta$, and one at the incorrect, negative value. Because the chain swaps between the two peaks, there is significant weight at $\beta=0$, and this signal is not detectable as non-GR. For all injections, the constituent masses were  $m_1/m_2 = 1.42/2.0 M_\odot$.}
\end{figure}

The overall behavior in Fig.~\ref{BFtheta} is as expected. For low $f^*$, the signal is clearly non-GR, and when $f^*$ is very high, GR is favored. There is an unexpected region in the middle, however, in which the BF grows with $f^*$. We can further investigate this region by looking at the posterior distributions for the ppE strength parameter, $\beta$, plotted in Fig.~\ref{fourPDF}. At first, there is only one peak in the posterior, and it is centered at the correct, injected value of $\beta$. As $f^*$ becomes larger, a secondary maximum appears, centered at an incorrect value of $\beta$. For some values of $f^*$, these two maxima fit the data equally well, which leads to significant posterior weight at $\beta = 0$, leading to a BF that favors GR. As $f^*$ grows, the secondary maximum becomes a better fit to the data, until eventually the GR value of $\beta = 0$ wins out. This can be understood by noting the relationship between the BFs and the aLIGO noise curve we have used. The BF is largest when $f^*$ is in the region of highest sensitivity for the detector.

\begin{figure}[H]
\begin{center}
\begin{tabular}{cc}
\hspace*{-8mm} \includegraphics[width =9.5 cm]{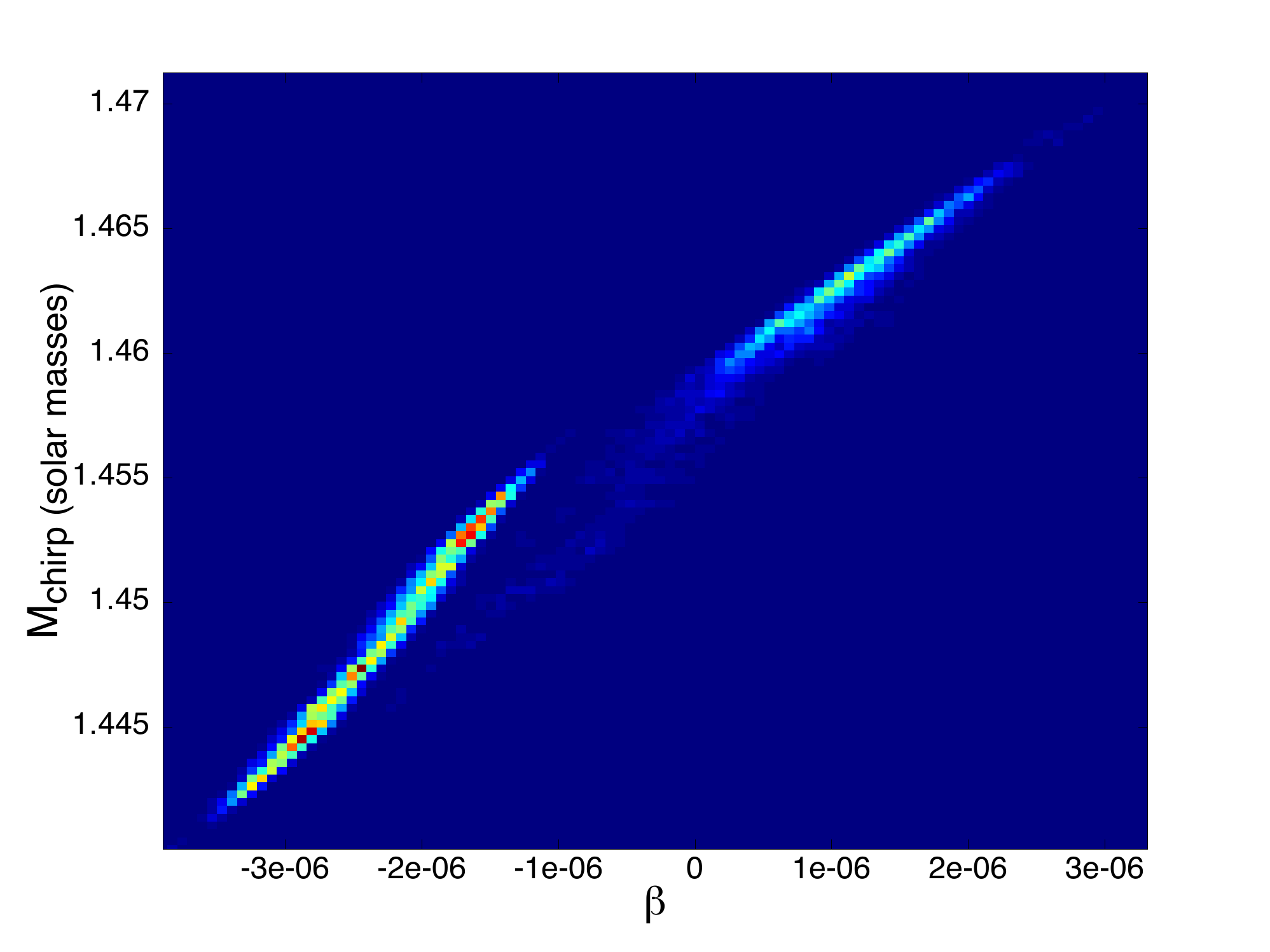} 
\end{tabular}
\end{center}
\vspace*{-0.2in}
\caption{\label{betaMchirp} The correlation between $\beta$ and $\mathcal{M}$, generated from a signal of the form in Eq.~(\ref{eq:phasetheta}), with $f^* = 47.5$. The two separate maxima in the likelihood are clearly visible, as well as the strong correlation between these two parameters. The injected chirp mass was $\mathcal{M} = 1.463 M_\odot$, with constituent masses  $m_1/m_2 = 1.42/2.0 M_\odot$.}
\end{figure}

This behavior can be understood further by examining the correlations between $\beta$ and the other source parameters, for example in Fig.~\ref{betaMchirp}. This plot shows the two-dimensional posterior distribution for $\beta$ and the chirp mass, $\mathcal{M}$. In this example, there is clear correlation visible between the two parameters, and two peaks in the posterior are clearly visible. While these two peaks both represent good fits to the data, there is also significant weight between them, which means significant weight at $\beta=0$, and thus a BF that favors GR.

In addition to testing the standard ppE templates, we also use templates of the form in Eq.~(\ref{eq:phasetheta}), in which we allow $f^*$ to be a parameter that is determined by the data. We use these templates to recover the same signals injected in the previous study, and again use MCMC techniques to calculate the BF between GR and non-GR models. The results are also plotted in Fig.~\ref{BFtheta}. These enhanced templates show the same qualitative behavior as the standard ppE templates, although their overall performance is better. This is to be expected, as this template family can perfectly match the injected signals.

\begin{figure}
\begin{center}
\begin{tabular}{cc}
\hspace*{-8mm} \includegraphics[width =9.5 cm]{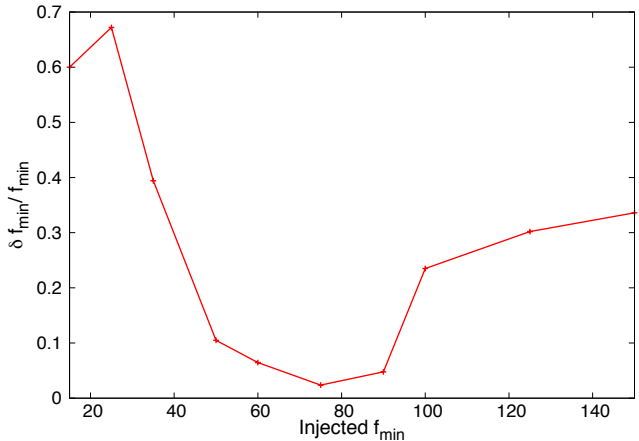} 
\end{tabular}
\end{center}
\vspace*{-0.4in}
\caption{\label{deltaF} Fractional uncertainty in the recovered value of $f^*$, for different injected values of $f^*$. The uncertainty is inversely proportional to the BF in favor of the non-GR model -- i.e.~when the BF indicates a clear departure from GR, the $f^*$ parameter is recovered with high accuracy. Again, the injected signals had masses  $m_1/m_2 = 1.42/2.0 M_\odot$.}
\end{figure}

An additional point of interest for the enhanced templates is the precision with which we are able to recover the injected parameter, $f^*$. The uncertainty in the recovered $f^*$ is plotted as a function of the injected $f^*$ in Fig.~\ref{deltaF}. The precision with which we can measure this parameter depends on the BF, as expected. Even so, for a large range of values, we are able to recover this parameter quite accurately. If a signal of this type were detected in the data, information about $f^*$ would be useful for theorists attempting to learn about the underlying gravitational theory.

\section{Inspiral-Merger-Ringdown Signals}
\label{sec:ins-analysis}

 In this section, we analyze some of the issues that can arise from using inspiral-only templates in tests of GR, as well as some of the science we can do by including merger and ringdown in our analysis. We explore the problem of testing GR using full GW injections - i.e., injected waveforms that include merger and ringdown in addition to inspiral. In the first subsection, we investigate what can happen when we use inspiral-only templates to extract full signals. In the second subsection, we consider a family of ppE templates that includes non-GR parameterizations of the merger and ringdown stages. 

\subsection{Extracting with Inspiral-only ppE Templates}

Typically, GWs from binary systems are talked about as if they have three discrete parts -- inspiral, merger, and ringdown (IMR). The inspiral is the part of the waveform that is generated while the two bodies are still widely separated, and thus slowly spiraling towards each other due to the emission of GWs. The merger is the most difficult part of the signal to model analytically, and is the portion in which the two bodies are very near each other and moving very quickly -- eventually becoming one object. Finally, the ringdown stage is produced after the two bodies have merged, as the resulting object relaxes to its final state. 

The inspiral portion of GW signals has been calculated in several alternative gravity theories~\cite{Yunes:2009ke,Berti:2004bd,Damour:1998jk,Berti:2005qd,Will:2004xi,Will:1997bb,Yagi:2013mbt,Yagi:2011xp} (for a recent review, see~\cite{Yunes:2013dva}). These calculations were then used as motivation for the inspiral waveforms in the ppE family. The merger phase, unfortunately, has not been calculated in any non-GR theories -- GWs outside of GR may even lack ringdowns altogether, and almost certainly would have different relationships between the system parameters and the quasi-normal modes~\cite{Berti:2009kk,Dreyer:2003bv,Gossan:2011ha}. Even within GR, the merger stage must be calculated numerically, and connected phenomenologically to the analytic solutions for inspiral and ringdown~\cite{Santamaria:2010yb,Ajith:2009bn}. This means that we lack theoretical motivation for non-GR merger and ringdown templates, and thus when testing GR we usually choose to use only the inspiral portion of the signal where we have an analytic expression for GR waveforms, and well-motivated templates for alternative gravity. 

We often discuss the three stages of GWs as if they are clearly separable, but the transition from inspiral to merger and merger to ringdown is a somewhat arbitrary distinction. One common choice is to take the transition from inspiral to merger to be the frequency of a test-particle at the innermost stable circular orbit (ISCO) of a Schwarzschild black hole, $f_\ISCO=1/(6^{3/2}\pi M)$. For full waveforms that include the merger and ringdown, the transition from inspiral to merger is smooth, and can begin to have effects earlier or later than this, depending on the system. When using inspiral-only templates, as is commonly done in GW data analysis~\cite{Colaboration:2011np,Abbott:2009qj,Abbott:2009tt}, the question of when (i.e.~at what frequency) to terminate the waveforms is not a trivial one.

In order to use the full, three-stage signal model for detection and characterization of GWs, we need an efficient template family that can capture the full signal. In this paper, we use PhenomC waveforms in our analyses of full IMR signals~\cite{Santamaria:2010yb,Ajith:2009bn}. In these waveforms, the inspiral stage is modeled within the PN approximation. The merger stage has been studied numerically~\cite{Hinder:2010vn,Hannam:2009rd,Baker:2005vv,Campanelli:2005dd,Pretorius:2005gq}, and is approximated analytically. The final stage, ringdown, is modeled from black hole perturbation theory analytically. These three pieces of the waveform are stitched together with matching procedures and calibrated against full numerical results to produce full, IMR waveforms.

Our goal is to determine where inspiral ends in a self-consistent way -- one that does not lead to significant biases in the recovered parameters. When using GR inspiral templates to fit a full GR IMR signal, the biases arise primarily in the recovered value of the total mass. This is because the cutoff frequency of the template is determined by the total mass, and so the inspiral waveform ``stretches'' in the frequency domain to fit some of the merger power by changing this parameter. Although this type of bias is clearly undesirable, the true danger arises when using ppE inspiral signals to fit the full waveform. In this case, not only the total mass, but the ppE strength parameter, $\beta$, change to attempt to fit some of the merger/ringdown power. This can lead to a recovered value of $\beta$ that is not consistent with GR, and thus the claim of a detection of a deviation from GR in the GW signal.

\begin{figure}[t]
\begin{center}
\begin{tabular}{cc}
\hspace*{-8mm} \includegraphics[width =9.5 cm]{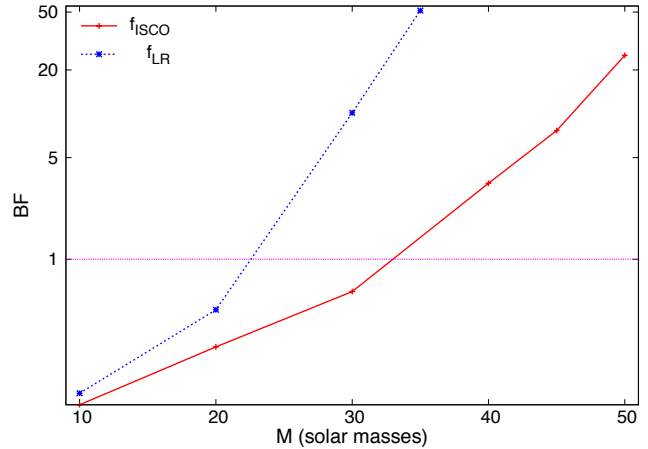} 
\end{tabular}
\end{center}
\vspace*{-0.4in}
\caption{\label{BF2} BFs between GR and ppE templates. The injected signals were GR, PhenomC waveforms, and they were recovered using inspiral-only ppE waveforms. The dashed (blue) line shows the BFs calculated by using the frequency at the light ring as as the cutoff frequency for the waveforms. The solid (red) line shows BFs calculated by using $f_\ISCO$ as the cutoff frequency. A BF larger than 1 indicates a preference for the non-GR model.}
\end{figure}

\begin{figure}[t]
\begin{center}
\begin{tabular}{cc}
\hspace*{-8mm} \includegraphics[width =9.5 cm]{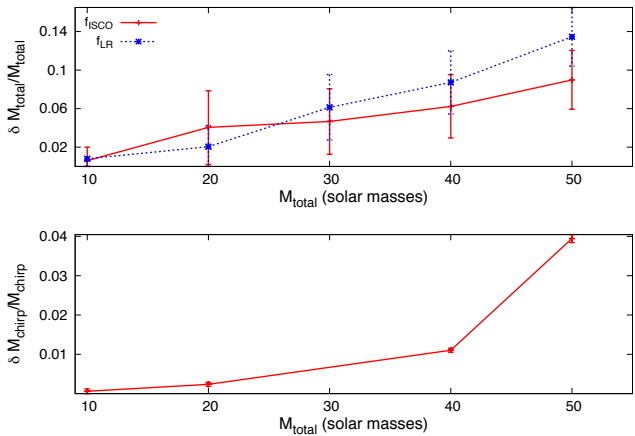} 
\end{tabular}
\end{center}
\vspace*{-0.4in}
\caption{\label{MTshift} (upper panel) The bias in total mass, $M$, recovered when using an inspiral-only, GR signal to fit an IMR, GR signal. The dashed (blue) line was calculated using the light ring to determine cutoff frequency, and the solid (red) line used the ISCO. The error bars show the 1-$\sigma$ limits for the recovered values. For high-mass systems, the bias nears $10\%$. For lower-mass systems, the recovered mass is very close to the injected value. Each injected signal had SNR $\sim$ 25. (lower panel) The bias in chirp mass, $\mathcal{M}$, recovered for the same systems injected in the upper panel, with the cutoff frequency determined using the ISCO. This plot illustrates that the recovered value of $\mathcal{M}$ is not strongly affected by the presence of merger and ringdown in the signal.}
\end{figure}

Figures~\ref{BF2} and \ref{MTshift} show some of the consequences that can arise when using inspiral-only templates to analyze a signal that includes merger and ringdown. To generate both figures, we injected PhenomC waveforms, with total mass ($M = m_1 + m_2$) beginning at $M = 10 M_\odot$, up to $M = 50 M_\odot$. We then recovered these signals using inspiral-only GR templates, as well as inspiral-only ppE templates with $b=-2$, which corresponds to a 1.5 PN correction to the GW phase. We used two different determinations of the cutoff frequency for the inspiral waveforms. In one set of runs, we used $f_\IM = f_\LR$, where $f_\LR$ is the frequency of a test-particle at the light ring of a Schwarzschild black hole. For the other set, we used $f_\IM = f_\ISCO$, the ISCO frequency. The choice of cut-off frequency had a significant effect on our results, signaling a departure from GR for sufficiently massive systems, even when the injection had none. In all cases, we again use the neutron-star-binary optimized aLIGO noise curve.

In Fig.~\ref{BF2}, we plot the BF of ppE vs. GR templates recovered from these signals. When the BF is larger than $1$, the non-GR model is preferred. Even though all injections are GR signals, the BF shows a preference for non-GR models for $M >  30 M_\odot$ when the light ring frequency is used to terminate the waveforms, and for $M > 40 M_\odot$ when the transition frequency for the waveforms is set to $f_\ISCO$. If we are not careful to use inspiral-only templates only for low-mass systems, then, we could mistakenly claim the detection of a deviation from GR.

This growth in BF in favor of non-GR models is accompanied by a growing bias in the recovered value for $M$. This is illustrated in Fig.~\ref{MTshift}, where we plot $\delta M/M$, where $\delta M$ is the difference between the recovered best-fit value for $M$ and the injected value. For the systems we analyzed, the discrepancy between the recovered and injected masses was never larger than $8\%$, even though, as expected, the templates using a cutoff frequency different from the injected transition frequency performed worse than the others.

It is not surprising that analyses using inspiral-only templates are only dependable for low-mass systems. As indicated above, the transition from the inspiral portion of the waveform to the merger portion depends upon the total mass of the system in question. This means that for a low-mass system, most of the SNR of the signal is contained in the inspiral, whereas for a high-mass system, a large fraction is in the merger and ringdown. Table \ref{SNRpercent} lists the inspiral/merger transition frequency and the fraction of SNR contained in the inspiral for systems of varying total masses.

\begin{table*}[ht]
\begin{center}
    \begin{tabular}{c|c|c|c|c }
    \hline\hline
    Total Mass ($M_\odot$) &$ f_\ISCO$ (Hz) &$\%$ SNR before ISCO&$f_{10\rm{M}}$ (Hz)&$\%$ SNR before $r=10\rm{M}$  \\ 
    \hline
    10&879&100&704&98\\
    20&220&93&393&47\\
    30&147&82&254&24\\
    40&110&72&188&12\\
    50&88&61&150&7\\
  \hline\hline
           \end{tabular}
   
    \caption{ \label{SNRpercent}ISCO frequency, frequency at $r=10\rm{M}$, and percentage of total SNR accumulated before these two frequencies for systems of different total mass.}
\end{center}
\end{table*}

The fact that using inspiral-only templates to fit IMR signals will lead to parameter biases has been understood for some time~\cite{Lindblom:2010mh,Lindblom:2009ux,Bose:2008ix,Cutler:2007mi,Pan:2011gk,PhysRevD.63.082005}, and to this point the method for avoiding these biases has been to use this type of signal only when analyzing low-mass systems. Here we present a simple, two-stage technique that allows us to use some higher mass systems in inspiral-only analyses:

\begin{itemize}
\item Analysis I: run a standard inspiral-only template analysis on the full signal, using the self-consistently determined frequency $f_\ISCO$ as a cut-off.
\item Analysis II: using the (biased) value of $M$ recovered in Analysis I, low-pass filter the data to remove everything above a frequency corresponding to $r = 10 M$, where
$r$ is the separation distance of the binary, and re-run the analysis.
\end{itemize}

We also present a separate method, Analysis III, in which the full IMR signal is fit using a two-part inspiral template. This template consists of two inspiral templates with independent mass parameters - one to fit the low-frequency part of the signal, and one to fit the high-frequency part. That is,
\be
\tilde{h}(f) = \begin{cases} \tilde{h}(f, \vec{\lambda}, M_1, \mathcal{M}_1, \beta_1) & \mbox{if} \,\,\,f < f_{10M_1} \\	
					\tilde{h}(f, \vec{\lambda}, M_2, \mathcal{M}_2, \beta_2)  & \mbox{if} \,\,\,f_{10M_1} \le f < f_{LR M_2},
					\end{cases}
\ee
where $M_i$,$\mathcal{M}_i$, and $\beta_i$ indicate the total mass, chirp mass, and ppE strength parameter for either the early (1) or late (2) portion of the signal. The other parameters, represented by $\vec{\lambda}$, are the same in both portions of the template. The cutoff frequency for the early inspiral template is set to be the frequency corresponding to $r = 10 M_1$, and the cutoff for the late portion is set to be the lightring frequency corresponding to $M_2$. We know that the inspiral template should fit the early part of the signal with no biases in the recovered parameters, and that the late part will be biased because of the presence of merger and ringdown in the data. This two-part template allows the higher frequency portions of the signal to be fit with biased parameters without polluting the recovery of the true system parameters using the early portion of the signal.

\begin{figure}[]
\begin{center}
\begin{tabular}{cc}
\hspace*{-8mm} \includegraphics[width =9.5 cm]{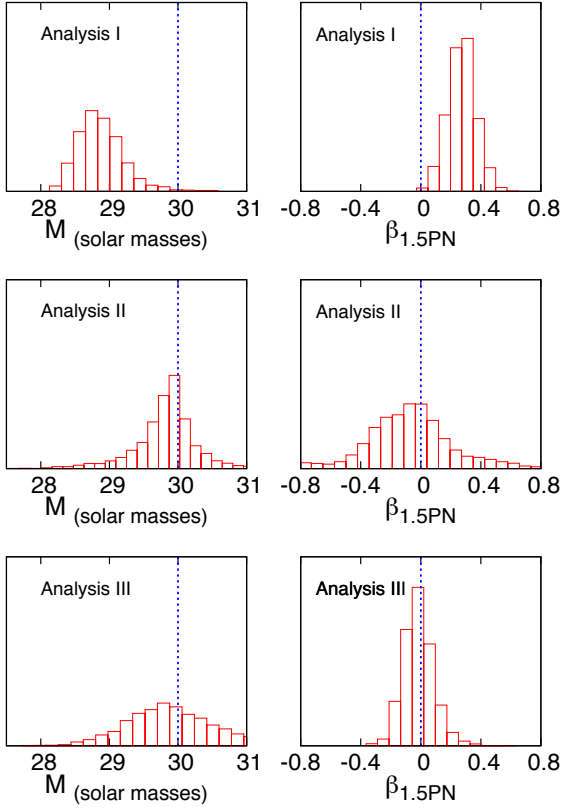} 
\end{tabular}
\end{center}
\vspace*{-0.4in}
\caption{\label{unbiasM30} The posterior distributions for M (left panels) and $\beta$ (right panels) for a 30 $M_\odot$ system, from Analysis I , II, and III. The bias in recovery of M is removed in Analyses II and III, as is the model preference for ppE over GR. The injected value is shown by the vertical line in each panel.}
\end{figure}
The results of these analyses are shown in Fig.~\ref{unbiasM30}. In this Figure, we plot the posterior distribution for $M$ and $\beta$ for Analysis I, Analysis II,  and Analysis III, all for the case where $M=30 M_\odot$. In Analysis II, by using the conservative cutoff frequency associated with $r=10M$ in our analysis, we are able to remove the bias in the recovered total mass, as well as avoiding making false claims that the signals are not consistent with GR. Similarly, in Analysis III, by allowing the template to fit only the low-frequency portion of the signal, we are also able to remove the biases in parameter recovery. One more feature to note is that the distributions recovered in Analyses II and III are broader than those in Analysis I -- this is because neither of these analyses fit the full signal, and so the full SNR is not available for these templates. Because of this effect, this type of procedure is only useful for high total mass systems that also have a high SNR. 

\begin{figure}[]
\begin{center}
\begin{tabular}{cc}
\hspace*{-8mm} \includegraphics[width =9.5 cm]{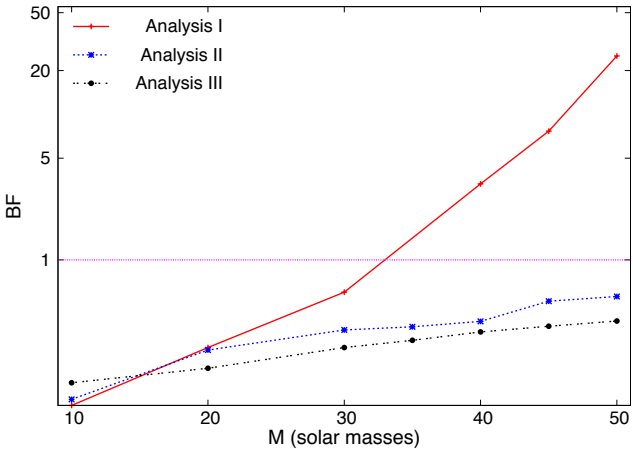} 
\end{tabular}
\end{center}
\vspace*{-0.4in}
\caption{\label{BF10M} BFs between ppE and GR templates, from Analysis I (solid/red), using $f_{\ISCO}$ as the cutoff frequency, and from Analysis II (dashed/blue), using $r=10M$ to calculate the cutoff frequency. All signals were GR signals. In Analyses II and III, the model selection process always favors GR.}
\end{figure}

The BF calculations are further illustrated in Fig.~\ref{BF10M}, where we show the BF between ppE and GR for GR injections of varying $M$, first for Analysis I using $f_\ISCO$ as the cutoff frequency, then for Analysis II, using $r=10M$, and finally using the two-part template of Analysis III. The more conservative Analysis II never results in the erroneous favoring of a non-GR model, and the two-part inspiral template of Analysis III performs even better.

\subsection{Extracting with Inspiral-Merger-Ringdown ppE Templates}

Because it is not possible to analyze systems of all masses with inspiral-only templates, the next problem we investigate is in using the full IMR waveforms to test GR. As we have mentioned, there are currently no concrete examples of merger/ringdown waveforms in non-GR gravity theories. This unfortunately means that we do not have strong theoretical motivations for what non-GR templates should look like for these parts of the signal. We do know, however, that by adding some flexibility to GR templates, via introducing parameters to the merger and ringdown stages, we will be able to fit a wider class of signals than with GR templates alone. We cannot at present know if the flexibility is enough to fit all possible non-GR signals, but if the extra parameters are recovered at their GR values, we can at the very least say that the data is consistent with GR.

This was exactly the philosophy followed in Ref.~\cite{Yunes:2009ke} when proposing the ppE template family. That paper, in fact, proposed a variety of families, including an IMR one. Restricting attention to a simplified version of such ppE IMR family~\cite{Yunes:2009ke}, we will consider the following templates:
\be
\tilde{h}(f) = h(f) = \begin{cases} h_\GR ( \exp{i\beta u^{b}}), & \mbox{if}\,\,\, f<f_{IM} \\ A_M f^{-2/3} \exp{i\delta}, & \mbox{if}\,\,\, f_{IM} \le f <f_{MR} 
 \\ \frac{A_R}{1+4\pi^2 \tau^2 (f-f_{MR})^2} & \mbox{if}\,\,\, f_{MR} \le f\end{cases},
 \label{IMRwave}
\ee
The inspiral portion is a standard PN inspiral with the inclusion of a single ppE phase term; we here restrict attention to the $b=-2$ case. The functional form of the merger is based on an analytic fit to numerical data, and includes two matching parameters, $A_M$ and $\delta$, to ensure continuity at the transition point. The ringdown is a single quasi-normal mode, with matching parameter $A_{R}$ to ensure continuity, real frequency $f_{MR}$ and decay time $\tau$. In GR, the decay time can be modeled via~\cite{Yunes:2009ke} 
\be
\frac{1}{\tau^{\GR}} = \frac{0.51 \eta^2 + 0.077 \eta +0.022}{\pi M}, 
\ee
while the transition frequencies between inspiral/merger and merger/ringdown can be modeled as~\cite{Yunes:2009ke} 
\begin{align}
f_{IM}^{\GR} & =\frac{ 0.29 \eta^2 +0.045 \eta+0.096}{\pi M} \nonumber \\
f_{MR}^{\GR} & =\frac{0.054\eta^2 +0.09\eta+0.19}{\pi M}.
\end{align}

\begin{figure}[t]
\begin{center}
\begin{tabular}{cc}
\hspace*{-8mm} \includegraphics[width =9.5 cm]{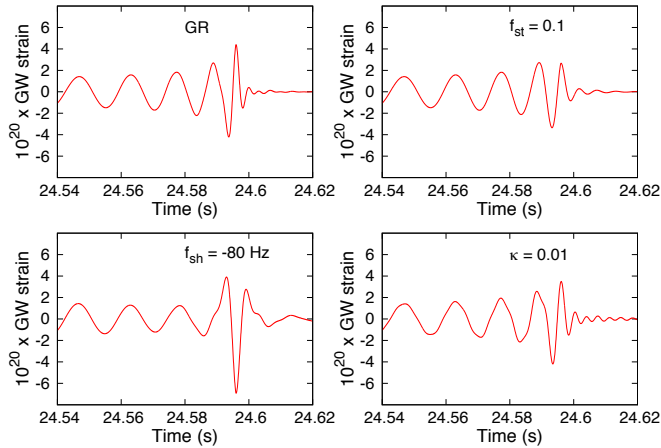} 
\end{tabular}
\end{center}
\vspace*{-0.4in}
\caption{\label{IMRwaves} Time-domain waveforms generated using the parameterization in Eq.~(\ref{IMRwave}), for an SNR 30 signal with total mass $M=50M_\odot$. Top left: GR waveform. Top right: $f_{\fstretch} = 0.1$: the merger portion of the waveform is compressed, but the frequency at which merger begins and the structure of the ringdown are unaffected. Bottom left: $f_{\shift} = -80 \; {\rm{Hz}}$: the beginning of merger is shifted to a lower frequency by $80$ Hz, but the duration of merger and the ringdown structure are unaffected. Bottom right: $\kappa = 0.01$: merger is unaffected, but ringdown is changed such that the decay is much slower than in GR.}
\end{figure}

To model deviations away from the GR expectation, we include four parameters that encode non-GR effects. These are $\beta$, the usual ppE phase parameter from the inspiral portion, and the following three new non-GR parameters:
\begin{itemize}
\item $f_{\shift}$ shifts the beginning of merger:  \\ \quad$f_{IM} = f_{IM}^{\GR} + f_{\shift}$
\item $f_{\fstretch}$ stretches the merger:  \\ \quad$f_{MR} = (f_{MR}^{\GR}-f_{IM}^{\GR}) f_{\fstretch} + f_{IM}$
\item $\kappa$ adjusts the value of $\tau$ from its GR value: \\ \quad$\tau^2 = \tau^2_{\GR}/\kappa$ 
\end{itemize}
The effects of these new parameters on the time-domain waveforms are illustrated in Fig.~\ref{IMRwaves}. GR is recovered when $(\beta,f_{\shift}, f_{\fstretch}, \kappa) = (0,0,1,1)$.

\begin{figure}[t]
\begin{center}
\begin{tabular}{cc}
\hspace*{-8mm} \includegraphics[width =9.5 cm]{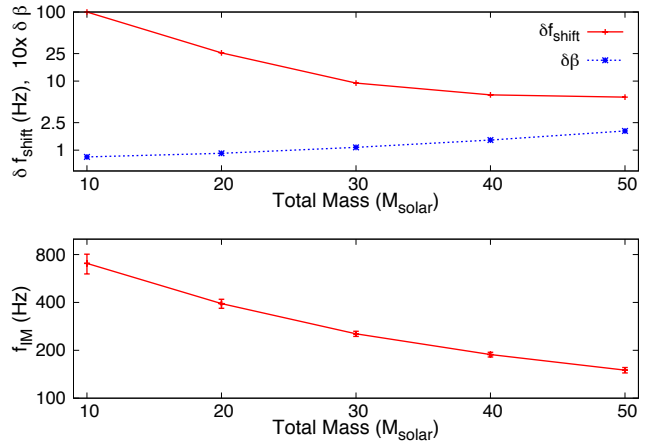} 
\end{tabular}
\end{center}
\vspace*{-0.4in}
\caption{\label{deltafshift} (upper panel) Uncertainty in the recovered value of $f_\shift$ (solid/red) and $10\times$ the uncertainty in the recovered value of $\beta$ (dashed/blue), for different injected values of $M$. The uncertainty in $f_\shift$ decreases as the total mass increases and the merger-ringdown portion of the waveform becomes more important. The uncertainty in $\beta$ increases due to correlations between the two parameters. All injected waveforms were GR signals. (lower panel) The injected value of $f_\IM$ (the transition between merger and ringdown) for the different injected systems. The error bars on this value are the uncertainty in $f_\shift$, as this parameter moves the transition between merger and ringdown.}
\end{figure}

\begin{figure}[]
\begin{center}
\begin{tabular}{cc}
\hspace*{-8mm} \includegraphics[width =9.5 cm]{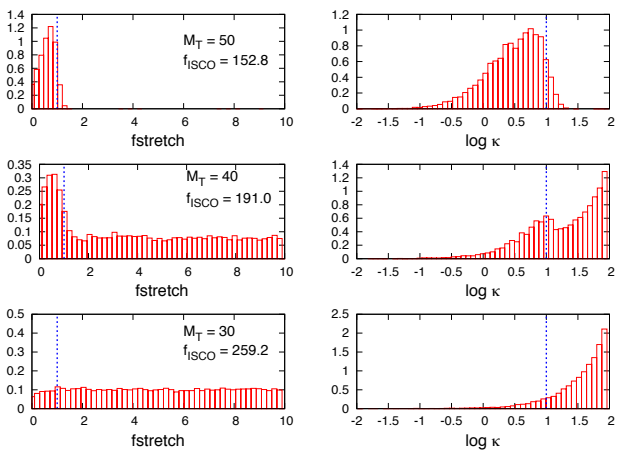} 
\end{tabular}
\end{center}
\vspace*{-0.4in}
\caption{\label{taustretch} Posterior distributions for the parameters $\kappa$ and $f_\fstretch$ for various values of $M$. As the total mass increases, the parameters go from being completely unconstrained to well-measured by the data. All injected signals were GR signals. The vertical line in each panel indicates the injected, GR value for that parameter.}
\end{figure}

In analogy with our previous work~\cite{Cornish:2011ys}, we assess how well these templates could be used to test GR by determining the range of values for each parameter that are consistent with a GR signal. We do this by injecting a GR signal of the form of Eq.~(\ref{IMRwave}) with  $(\beta,f_{\shift}, f_{\fstretch}, \kappa) = (0,0,1,1)$, in this case with SNR $\sim$ 25, and running an MCMC analysis with templates also of the form of Eq.~\eqref{IMRwave} but with free $(\beta,f_{\shift}, f_{\fstretch}, \kappa)$ to produce posterior distributions for these parameters. All standard GR parameters are also allowed to vary during these studies. The results we obtain are shown in Figs.~\ref{deltafshift} and \ref{taustretch}.

\begin{figure}[t]
\begin{center}
\begin{tabular}{cc}
\hspace*{-8mm} \includegraphics[width =9.5 cm]{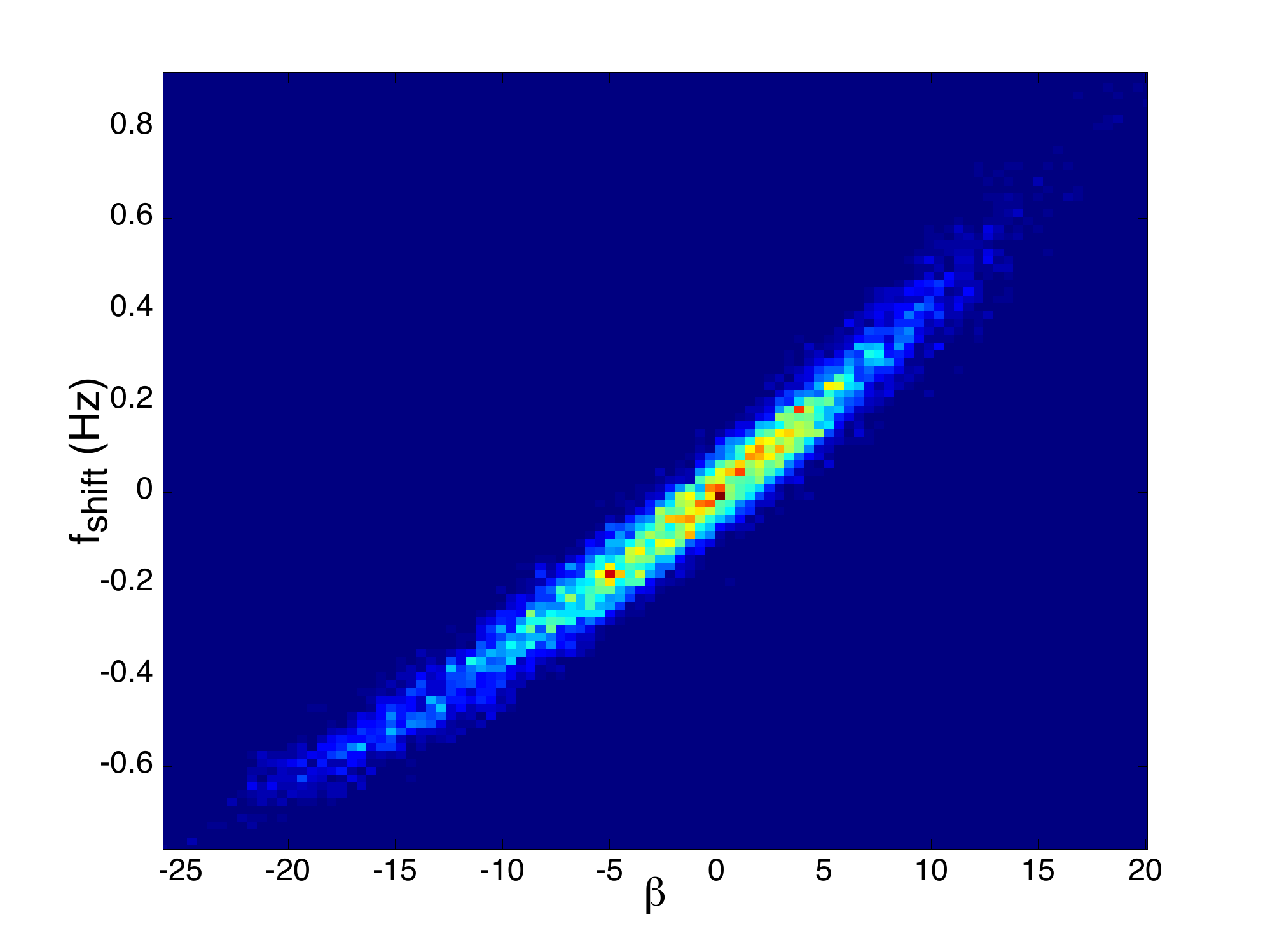} 
\end{tabular}
\end{center}
\vspace*{-0.4in}
\caption{\label{betashift} Correlation between the ppE phase parameter $\beta$, and the parameter $f_\shift$, that controls the start of the merger phase. This correlation is only present for systems with large M, and thus leads to an increase in the uncertainty in the recovered value of $\beta$ for these systems.}
\end{figure}

In Fig.~\ref{deltafshift}, we plot the uncertainty in the recovered values for $f_\shift$ and $\beta$, as a function of $M$. If we had detected a signal of the form injected, then our results indicate we would have been able to exclude the region $f_\shift$ and $\beta$ space above the curves shown in Fig.~\ref{deltafshift}. As expected, with increasing total mass the transition between inspiral and merger has a larger effect on the signal, and so $f_\shift$ is better constrained. The uncertainty in $\beta$, on the other hand, grows with increasing M. This can be understood by examining the correlation between these two parameters. As seen in Fig.~\ref{betashift}, which shows the two-dimensional posterior distribution for $\beta$ and $f_\shift$ for a system of total mass 50 $M_\odot$, the correlation is high.

To further illustrate what it means to constrain $f_\shift$, Fig.~\ref{deltafshift} also shows the frequency of transition between merger and ringdown for the injected signals. The error bars in this plot are the uncertainty in $f_\shift$, as this parameter moves the transition between merger and ringdown. When this transition occurs at a very high frequency, it is poorly constrained.

\begin{figure}[]
\begin{center}
\begin{tabular}{cc}
\hspace*{-8mm} \includegraphics[width =9.5 cm]{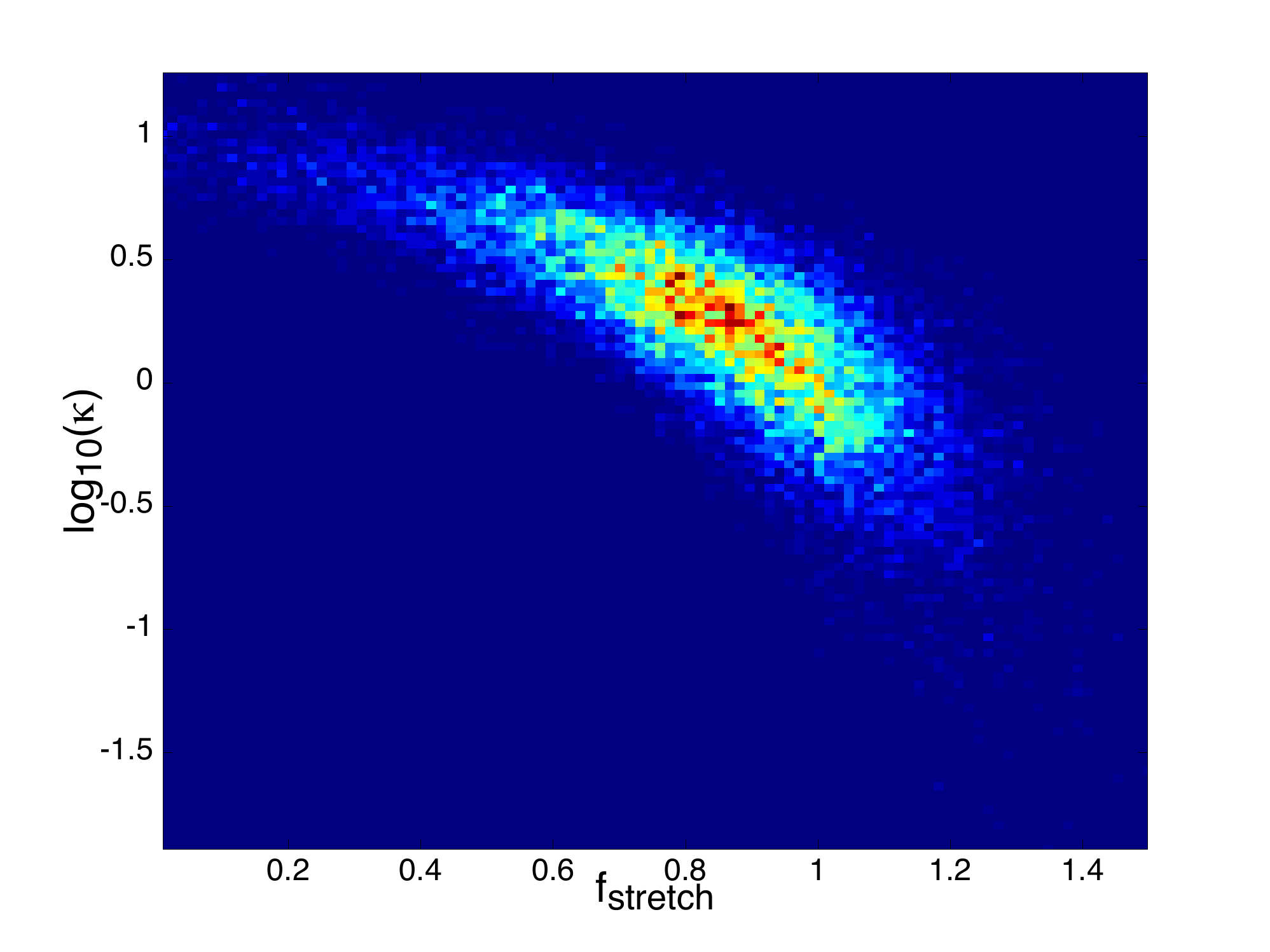} 
\end{tabular}
\end{center}
\vspace*{-0.4in}
\caption{\label{ltaustretch} Correlation between the parameter $\tau$, which affects the ringdown phase, and the parameter $f_\fstretch$, that controls the length of the merger phase.}
\end{figure}

For the other two parameters, $\kappa$ and $f_\fstretch$, the posteriors themselves are plotted in Fig.~\ref{taustretch}. We do so because, as can be easily seen in the figure, the posterior distributions for these parameters are highly non-Gaussian for low-mass systems, and so an estimate of the uncertainty is somewhat meaningless. The same general pattern is still apparent, however. For low-mass binaries, $\kappa$ and $f_\fstretch$ are essentially unconstrained within their prior ranges, which were both uniform. For $\kappa$, the distribution for the lowest mass system shows a clear preference for large values of the parameter. This is because a large $\kappa$ leads to a very short ringdown, and there is essentially no ringdown present in the actual data. Conversely, a small value of $\kappa$ leads to a long ringdown, and a correspondingly large amount of SNR contained in the ringdown.  As the mass of the system grows, the precision with which we could measure these two parameters (or analogously exclude non-GR deviations) increases. The peak of the distribution in each parameter, however, is not centered precisely on the GR value. This is because of correlation between the two parameters, which is shown in Fig.~\ref{ltaustretch} via the two-dimensional posterior distribution of $\kappa$ and $f_\fstretch$ at $M = 50 M_\odot$.

The systems for which we could use this parameterization to test GR are those for which the new parameters can be constrained. From Figs.~\ref{betashift} and \ref{taustretch}, we can conclude that: 
\begin{enumerate}
\item[(i)] Non-GR deviations to the merger and ringdown can be detected for total masses at or above $50 M_{\odot}$, 
\item[(ii)] Non-GR deviations to the inspiral become less detectable for larger total mass binaries. 
\item[(iii)] All non-GR parameters that characterize deviation to the merger and ringdown ($f_{\shift}$, $f_{\fstretch}$, and $\kappa$) can be constrained with IMR ppE templates. 
\end{enumerate}
These conclusions, of course, depend on the assumptions made in our analysis, chief among which are ${\rm SNR} \sim 25$, and neglecting spins and eccentricity. Including the latter, or studying signals with lower SNR will likely weaken the degree to which we can detect non-GR deviations. 

\section{Conclusion}
\label{sec:Conc}

Template-based searches for GWs are powerful tools that allow us to detect signals with low SNR and characterize the physical parameters of the sources. They are also an excellent means of testing GR. However, because templates impose strong prior beliefs on the form of the signals we expect to see, they can lead to strong biases in the analyses. This leads to a host of pitfalls that we must be aware of.

In this paper, we have explored several possible biases caused by using incorrect templates. First, we showed that there are possible departures from GR, not yet ruled out by experiment, that are so poorly matched by GR templates that they could evade detection by the searches performed to date. If these signals were detected using GR templates, there would by large biases in the recovered parameters. Using simple one-parameter ppE templates for detection ameliorates this problem significantly, but there are still a regions of parameter space in which, neither one-parameters ppE templates, nor GR templates would detect the signals.

We have also shown that the ppE templates can be used to detect deviations from GR that are very different in form from those they were created to capture. In particular, we showed that these ppE templates can capture modified gravity signals that ``turn on'' at a particular critical frequency. Furthermore, with a slight modification to the simplest ppE templates, deviations from GR generated by such ST theories could be measured in such a way that may help identify the parameters of these theories.

Additionally, we have shown that using inspiral-only templates is safe for low-mass binaries, but that stretching their use to higher masses would lead to biased parameter estimation, and could lead us to falsely believe we had discovered a GR deviation. The frequency we choose to use as a transition between inspiral and merger can have measurable effects on these results. Both of these issues can be avoided by using a simple, two-stage procedure for characterizing full signals with inspiral-only templates. Finally, we have explored  a simple IMR template family that contains a parameterization of deviations from GR that could possibly be used on full IMR signals. We found that such the merger-ringdown sector of ppE templates can be effectively constrained for signals of sufficiently high mass, where the merger-ringdown contributes significantly to the total SNR. Analogously, GR deviations that arise only in the merger and ringdown will be distinguishable from GR only for signals of sufficiently high SNR and high total mass.    

%%%%%%%%%%%%%%%%%%%%%%%%%%%%%%%%%%
%%%%%%%%%%%%%%%%%%%%%%%%%%%%%%%%%%
%%%%%%%%%%%%%%%%%%%%%%%%%%%%%%%%%%
\acknowledgments
L.~S. and N.~C. were supported by NSF grants PHY-1205993, PHY-1306702 and NASA grant NNX10AH15G. N. Y. acknowledges support from NSF grant PHY-1114374 and NASA grant NNX11AI49G, under sub-award 00001944, and also the NSF CAREER Award PHY-1250636.

%%%%%%%%%%%%%%%%%%%%%%%%%%%%%%%%%%
%%%%%%%%%%%%%%%%%%%%%%%%%%%%%%%%%%
%%%%%%%%%%%%%%%%%%%%%%%%%%%%%%%%%%
%%%%%%%%%%%%%%%%%%%%%%%%%%%%%%%%%%
\bibliography{oddsends}
\end{document}